\documentstyle[prd,aps,epsf,epsfig]{revtex} 
\begin{document}
\draft
\twocolumn[\hsize\textwidth\columnwidth\hsize\csname@twocolumnfalse\endcsname
%
\title{Percolation-like behavior of some optimal coalition formation models}
\author
{Z. N\'eda$^{1,*}$, R. Florian$^2$, M. Ravasz$^1$, A. Lib\'al$^1$ and G. Gy\"orgyi$^3$}
\address{
$^1$Department of Physics, Babe\c{s}-Bolyai University, RO-3400 Cluj, Romania\\
$^2$Center for Cognitive and Neural Studies, str. Saturn 24, RO-3400, Cluj, Romania\\
$^3$Department of Theoretical Physics, E\"otv\"os Lor\'and University, Budapest, Hungary }
\maketitle
\centerline{\small (Last revised \today)}

\begin{abstract}
The ground-state of an infinite-range Potts glass-type model
with $\pm J$ bonds and unrestricted number
of states is used to investigate coalition formation.
As a function of the $q$ probability
of $+J$ bonds in the system it is found that the
$r$ relative size of the largest cluster (a cluster being the
group of elements in the same state) shows a percolation like
behavior. By a simple renormalization approach and several optimization
methods we investigate the $r(q)$ curves for finite systems sizes.
Non-trivial consequences for social percolation problems are discussed.

\end{abstract}

\vspace{2pc}
]

\vspace{1cm}


\section{Introduction}

The Potts glass was originally introduced for studying various non-magnetic random
orientational \cite{glass1} and structural \cite{glass2} glasses, which do not
posses reflection or rotational symmetries. Apart of the specific solid-state and
statistical physics applications, the infinite-range (or mean-field) version of the
model recently received renewed interest from the view-point of coalition formation
phenomenon in sociological systems \cite{axelrod,galam,florian}.
From this perspective the primary interest is in the
ground-state of the infinite-range (or mean-field) $p$-state Potts glass.

The infinite-range p-state Potts glass is usually defined by the Hamiltonian:
\begin{equation}
H=-p \sum_{i<j} J_{ij} \delta_{\sigma(i) \sigma(j)}
\label{potts}
\end{equation}
The $\sigma(i)$ Potts states can take the values $0,1,2.....,p-1$. The sum is extended over all
$N(N-1)/2$ pairs in the lattice, $\delta_{mn}=1$ if $m=n$ and $\delta_{mn}=0$
otherwise. The $J_{ij}$ bonds are randomly distributed quenched variables
with $J_0/N$ mean, and the variance is presumed to scale as $N^{-1}$. The system has a non-trivial frustration and computing the
thermodynamic parameters is a complex task.
The above model has been extensively studied by many authors through different methods
\cite{elderfield,erzon,gross,kirckpatrick,cwilich,santes,yokota,peters,dillman}.
Within the replica theory a self-consistent description of the
low-temperature glassy phase was obtained \cite{gross,cwilich}. For $p>2$ and low
enough temperatures it was found
\cite{elderfield,santes} that the infinite-range Potts glass is finally always
ferromagnetic.
Here we consider a special
case of the infinite-range Potts glass, which
can be useful in understanding  some universalities for coalition formation phenomena
in sociological systems.
The main difference between the original Potts glass and the model studied here is that we
we consider an important class of the $J_{ij}$ bonds, where the variance
scales as $N^{-2}$. Also, we consider unrestricted number of $p$ Potts states ($p=N$),
and restrict the study on the ground-state ($T=0$). In the $N\rightarrow \infty$
limit an interesting percolation-like transition is then revealed which is studied
by different approximations for finite system sizes.

\section{The Model}

In order to describe the process of aggregation or
coalition-formation phenomena in politics, economics or sociological systems
we introduce a model similar to the original Potts glass model.
In such a system given a set of $N$ actors (in our case the Potts variables)
we define an associated distribution of bilateral propensities towards either
cooperation or conflict \cite{axelrod}. The actors might be countries which ally into international coalitions,
companies that adopt common standards and strategies, parties that make alliances,
individuals which form different interest groups, and so on. The propensities
will define the $Z_{ij}$ interactions between the actors. The $Z_{ij}$
bond is positive if there is a tendency towards
cooperation and negative if it is a conflicting  relation between actor $i$
and $j$.
For simplicity reasons let us assume first that the $Z_{ij}$ links are symmetric
($Z_{ij}=Z_{ji}$), however later the case without this assumption is also considered.
In addition to this, each actor has an $S_i>0$ weight-factor which characterizes its
importance or size in the society.  This may be a demographic, economic or military factor, or
an aggregate parameter. The  question then arises as what kind of coalitions are
formed in order to optimally satisfy the conflicting interactions.
In particular we are interested in the size of the largest cluster in the
optimal state.

This non-trivial optimization problem can be mathematically formulated in
the formalism of a zero-temperature Potts glass type model.  To prove this,
we define a cost-function, $K$, (a kind of energy of the system) that
is increasing with $S_i S_j \mid Z_{ij} \mid$ whenever two conflicting actors
($i$ and $j$) are in the same
coalition or two actors which have a tendency towards collaboration are in
different coalition. The cost-function is zero, when no propensity is
in conflict with the formed coalitions.
The number of possible coalitions is unrestricted (maximal possible
number is $N$), and we denote the coalition in which actor $i$ is
by $\sigma(i)$.
The cost function then writes as:
\begin{equation}
K=-\sum_{i<j} \delta_{\sigma(i) \sigma(j)} Z_{ij} S_i S_j
+\frac{1}{2} \sum_{i<j} (Z_{ij} S_i S_j + \mid Z_{ij} S_i S_j \mid),
\label{cost}
\end{equation}
It is immediate to realize that for a given distribution of the $Z_{ij}$ interactions and $S_i$ weight-factors
the second term in equation (\ref{cost}) is constant (independent of the formed coalitions).
Minimizing the $K$ cost function
is equivalent with finding the ground-state of the (\ref{potts}) Hamiltonian
with $p=N$.
Instead of $S_i S_j Z_{ij}$ we now introduce the $J_{ij}N$ notation.
If $Z_{ij}$ and $S_i$ are independent of $N$ we have that $<J_{ij}>$ scales as $N^{-1}$,
and we introduce the notation: $J_0=N<J_{ij}>$.
We consider now a somehow trivial but practically important and general case, when
the variance of $J_{ij}$ scales as $N^{-2}$. (As an immediate example for this scaling is the
simple case when $S_i=S_j=1$ and $Z_{ij}$ is $+1$ with a probability $q$ and $-1$ with a probability
$1-q$.) For this choice, the $N\rightarrow \infty$ thermodynamic limit becomes simple, since
the disorder in the system scales out. The infinite-range Potts glass becomes thus equivalent
with a simple mean-field Potts-model, with $J_0$ interactions between the elements.
While for $J_0>0$ the system has minimal cost function when all elements
are in the same coalition, for
$J_0<0$ in the ground-state each element has to be in a different coalition. As a function of $J_0$ a
transition is thus expected. This transition resembles the one obtained in percolation or random
graph models. Since the temperature has no role in this phenomenon, we call
it geometrical phase transition.
In the present paper we study this geometrical phase transition
for finite $N$ values and simple $J_{ij}$ distributions.
The finite $N$ limit is however not as simple as the thermodynamic limit.
Frustration effects are important and finding the ground-state is a complex
NP hard optimization problem. (It is believed that for large $N$ the number of steps
necessary for an algorithm to find an exact optimum must, in general, grow faster than
any polynomial in $N$.) Several methods were used to investigate finite-size behavior in
the expected transition. First a simple renormalization approach was considered.
For small systems (up to $N=10$) an exact
enumeration was then used. For larger systems (up to $N=60$) Monte-Carlo type simulated
annealing and the recently proposed extreme optimization was applied.

The order parameter considered by us is the $r$ relative size of the largest cluster.
In the thermodynamic limit $r$ has the right behavior, for $J_0<0$  we get $r=0$, and for
$J_0>0$ we obtain $r=1$. More precisely, $r$ is computed as
\begin{equation}
r(J_0)= < \overline{ max_{(i)} \left \{ \frac{C_x(i,J_0))}{N} \right \} } >_x,
\label{op}
\end{equation}
where $C_x(i,J_0)$ stands for the number of elements in state $i$
for an $x$ realization of the $J_{ij}$ distribution, when \\ $<J_{ij}>=J_0/N$.
Since the ground-state might be degenerated (i.e. many
possible configurations with the same minimal energy might exist) we
make an average over all these states, denoted in (\ref{op}) by the over-line.
$<....>_x$ refers then for an ensemble average over $J_{ij}$.

We focus now on the simplest model in which we expect this transition, i.e. when
$J_{ij}$ is a two valued quenched random variable,
$J_{ij}=1/N$ with probability $q$ and $-1/N$ with probability $1-q$
(i.e. when $S_i=S_j=1$ and $Z_{ij}$ is $+1$ with a probability $q$ and $-1$ with a probability
$1-q$).
The distribution function of the $J_{ij}$ values writes as
\begin{equation}
P(J_{ij})=q \delta(J_{ij}-1/N)+(1-q) \delta(J_{ij}+1/N),
\label{prob}
\end{equation}
where $\delta(x)$ denotes the Dirac functional.
We assumed here that the $J_{ij}$ links are symmetric
($J_{ij}=J_{ji}$). It is immediate to realize that for this distribution:
\begin{eqnarray}
<J_{ij}>=(2q-1)/N,
(\Delta J_{ij})^2=\frac{4q(1-q)}{N^2}.
\end{eqnarray}
In the view of our previous arguments we expect that in the $N\rightarrow \infty$
limit the $r(q)$ curves will indicate a geometrical phase-transition at $q=1/2$.

\section{Renormalization approach}

Our elementary {\em renormalization} approach estimates in a mean-field manner the new
relative size of the largest state, whenever the system size is doubled. We start
from a system composed by only two elements (step 1). In the ground-state, the
probability to have these two elements in the same Potts state is $q_1=q$. The
relative size of the largest cluster is then $r_1=q_1+(1-q_1)/2$,
since the largest cluster will be the total system with probability $q_1$, and the
original half with probability $1-q_1$. In step 2 we now double the system size
by linking through all possible $J_{ij}$ connections two previous configurations
(A and B) with maximal relative size $r_1$, each of them having two elements.
Then, we reduce the
four $J_{ij}$ connections between the elements of A and B to a single one, and
transform the system into a configuration similar to the one from step 1. This
procedure is summarized in Fig.~1.

\begin{figure}[htp]
\epsfig{figure=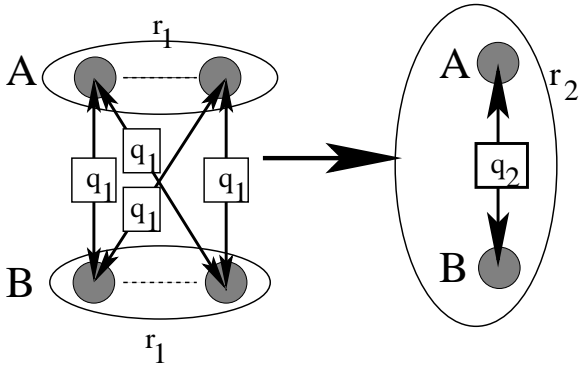,width=3.0in,angle=-0}
\caption{Schematics of the renormalization approach.}
\label{fig1}
\end{figure}

The new link will be positive ($+1$) with probability $q_2=q_1^4+
4q_1^3(1-q_1)+3q_1^2(1-q_1)^2$, and the new relative size of the largest
state is $r_2=q_2+(1-q_2)r_1/2$. The factor $3$ from the last term in the expression of
$q_2$ results by considering the new link positive with $1/2$ probability, whenever
there are two positive and two negative links ($6$ possible realizations in total).
This doubling procedure is then recursively repeated, leading to the simple
renormalization equations:
\begin{eqnarray}
q_{k+1}=q_k^4+4q_k^3(1-q_k)+3q_k^2(1-q_k)^2,
\label{renq} \\
r_{k+1}=q_{k+1}+(1-q_{k+1})\frac{r_k}{2}.
\label{renr}
\end{eqnarray}
The size of the system after $k$ steps is $N=2^k$.

On the $[0,1]$ interval, iteration (\ref{renq}) has two stable fix-points: $0$ and $1$.
There is also an unstable fix-point $q=1/2$. Starting the iteration from
$q \in [0,1/2)$ we get $lim_{k\rightarrow \infty} q_k=0$ and $lim_{k\rightarrow \infty} r_k=0$.
Choosing $q \in (1/2,1]$ we get $lim_{k\rightarrow \infty} q_k=1$ and $lim_{k\rightarrow \infty}
r_k=1$. These results suggests that in an infinite system we have two distinct phases
separated by $q_0=1/2$, as expected. In phase I the $r$ order parameter converges to $0$, and
in phase II $r$ converges to $1$. We get thus the expected percolation-like
transition as a function of $q$.

\begin{figure}[htp]
\epsfig{figure=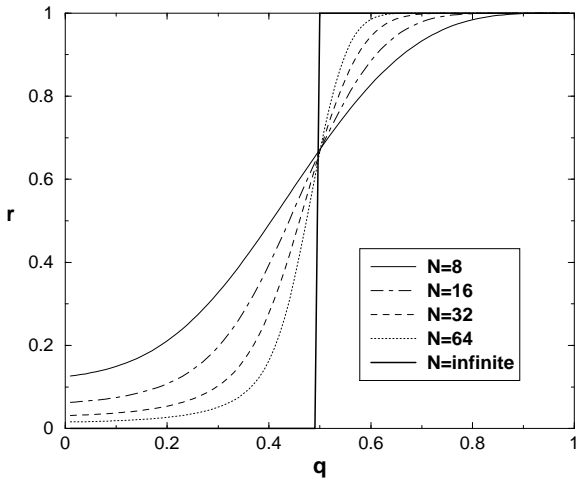, width=3.0in,angle=-0}
\caption{Renormalization results.}
\label{fig2}
\end{figure}

Using equations
(\ref{renq}-\ref{renr}) we can also easily plot the $r(q)$ curves for different system
sizes. Results in this sense are presented
in Fig.~2. These results support our previous arguments.

\section{Exact enumeration}

For small system sizes ($N\le 10$) exact enumeration is possible. This means that
one can computationally map the whole phase-space (all $\sigma(i)$ realizations)
for a generated $J_{ij}$ configuration and determine the minimum energy states.
Moreover, for $N\le 7$ it was also possible to map all $J_{ij}$ configurations
as well, our results up to $N=7$ are thus exact. In the $7<N\le 10$
interval, although the minimum energy states are exactly found, due to greatly
increased computational time and memory needed it was possible to generate only
a reasonable ensemble
average for $J_{ij}$ (5000 configurations). Results are plotted on Fig.~3.

\begin{figure}[htp]
\epsfig{figure=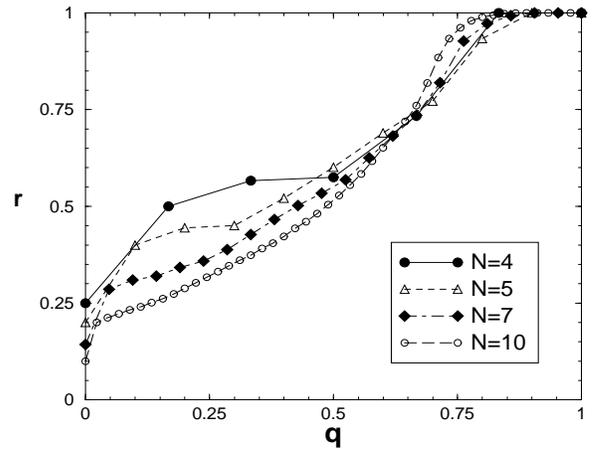,height=2.4in, width=3.0in,angle=-0}
\caption{Exact enumeration results for small systems. }
\label{fig3}
\end{figure}

We performed this exact enumeration with two purposes. First, we checked
the trends of the $r(q)$ curves as a function of increasing system size.
Secondly, these results offer a good "standard" for our less rigorous Monte-Carlo
type optimization methods, used for larger system sizes. As the results in Fig.~3
shows the $r(q)$ curves have a similar trend as those suggested by our renormalization
approach, i.e. as the system size increases we find increasing slopes for
$r(q)$ around a nontrivial $q$ value.

\begin{figure}[htp]
\epsfig{figure=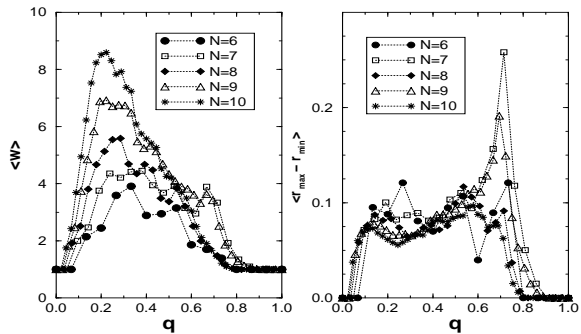,height=2in, width=3.0in,angle=-0}
\caption{Average degeneration level of the ground-state (a.) and
difference between the maximal and minimal $r$ value for different
coalition structures in the ground state (b.), both as a function
of the $q$ probability of $+$ interactions between the elements.}
\label{fig4}
\end{figure}

By exact enumeration we have also studied the degeneration level of the
ground state. For a given $J_{ij}$ bond-configuration,
many different coalition structure can have the same
ground-state energy. We can define thus a $w$ degeneration level
for each ground-state, and for a fixed $q$ value we can
calculate the $<w>$ ensemble average over all bond configurations.
Different coalition structures in the ground-state might be characterized
by different $r$ values, as well. For a given bond
configuration the difference between the
maximal $r$ value ($r_{max}$) and the minimal one ($r_{min}$) will
characterize the maximal possible deviation in the order parameter.
An ensemble average over this quantity ($<r_{max}-r_{min}$> will give
information about the differences which are possible to get in the
$r$ order parameter, while choosing another equally optimal
coalition structure. The values of $<w>$ and $<r_{max}-r_{min}$ have been
calculated as a function of the $q$ parameter. The obtained
results are presented in Fig.~4

\vspace{0.35in}

\section{Monte Carlo type optimization}

Monte-Carlo type optimizations were used for computing the ground-state of
larger systems. We considered both the classical simulated annealing
\cite{science} and the recently proposed extreme optimization method \cite{eo}.
Both approaches are rather time-consuming and the necessary computational time
increases sharply with system size. Our computational resources allowed to study
systems with sizes up to $N=60$.

Simulated annealing has been implemented in the standard fashion \cite{science}.
For the extreme optimization method we generalized the originally proposed method
\cite{eo} by considering a two-step algorithm. In the first step we performed the
usual optimization after the energies of the elements. As suggested in
\cite{eo} we assigned a given fitness to each Potts element and ranked all the
variables according to their fitness. Considering the $P(k) \sim k ^{-\tau}$
probability distribution over the rank, $k$, we then select an element for which the
state will be changed. For this first step we found the optimal value of $\tau=0.25$.
In the second step we decide the new state
of the chosen element by a similar procedure. For this second step the optimal
value of $\tau$ proved to be $4$.

Simulated annealing and extreme optimization gave identical and practically
indistinguishable results. Therefore in Fig.~5 we plot only the simulated
annealing results. The shape of the
\begin{equation}
\Delta r(q) = \sqrt{<\overline{r^2(q)}>_x-<\overline{r(q)}>_x^2}
\label{fluct}
\end{equation}
standard deviation was also computed (Fig.~4b),  suggesting  a non-trivial peak.
In Fig.~5 the curves for $N=10$, $20$, $30$ and $40$
were obtained with an ensemble average of $5000$ realizations, and the results
for $N=60$ with a statistics of $1000$ realizations. For $N=10$ the Monte-Carlo type
results are in perfect agreement with the ones from exact enumerations (Fig.~5a),
giving confidence in the used stochastic simulation methods.

\begin{figure}[htp]
\epsfig{figure=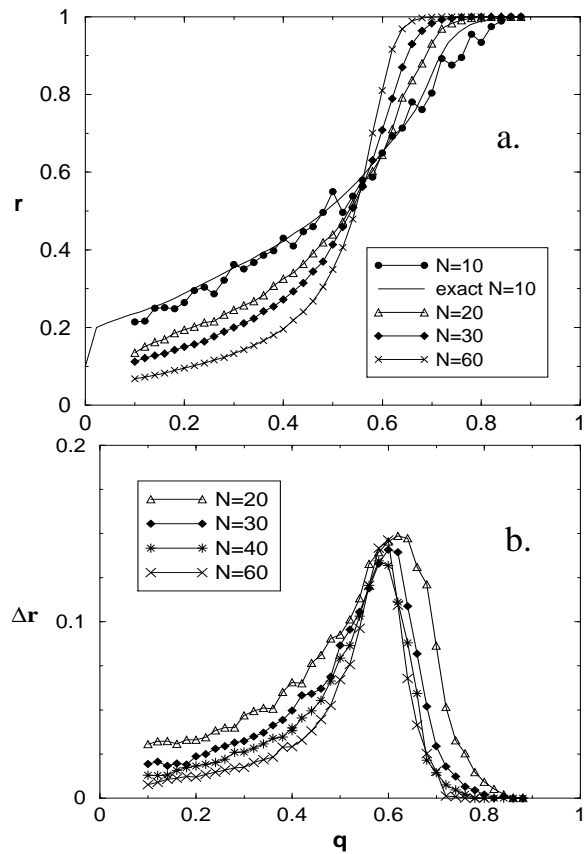,width=3.0in,angle=-0} \caption{Monte-Carlo
optimization results: (a.) variation of the order parameter, (b.)
standard deviation of the order parameter as a function of $q$.
For comparison purposes on Fig.~4a the exact enumeration results
for $N=10$ are also shown (continuous line).} \label{fig5}
\end{figure}

Our estimates suggest that extreme optimization was
faster by a factor of at least two, in comparison with simulated annealing.
However,
we found that extreme optimization is also strongly affected by the increasing system
size, and for $N>60$ we couldn't get any good statistics in reasonable computational time.

The results plotted in $Fig.~5$ support the expected geometrical phase
transition in the system.  As the system size increases the $r(q)$ curves show a more and more
sharper trend in the vicinity of $q=1/2$. Also, the $\Delta r(q)$ standard deviation
exhibits a non-trivial peak, which gets sharper and closer to
$q=1/2$ as the system size increases.
By extrapolating the obtained results as a function of $N$ for
$q=0.1$, $q=0.3$ and $q=0.7$, one can show that $r\rightarrow 0$ as a power-law
for $q=0.1$ and $q=0.3$, and $r\rightarrow 1$ for $q=0.7$ (Fig.~6).
This proves the
existence of the presumed phases.

Dropping the symmetry requirement for $J_{ij}$introduces an extra frustration
in the system. While for symmetric $J_{ij}$ only subsets with more than two elements
can be frustrated, in the asymmetric case subsets of two elements can become
already frustrated. It is interesting to note however, that the nature of the
observed transition is not affected by dropping this symmetry requirement
and again, the same geometrical phase transition should appear in $q_c=1/2$.
Up to $N=10$ we computed the $r(q)$ curves
by exact enumerations and for $N=20$ and $30$ we used the extreme optimization method.
No important deviations from the symmetric case were found.

\begin{figure}[htp]
\epsfig{figure=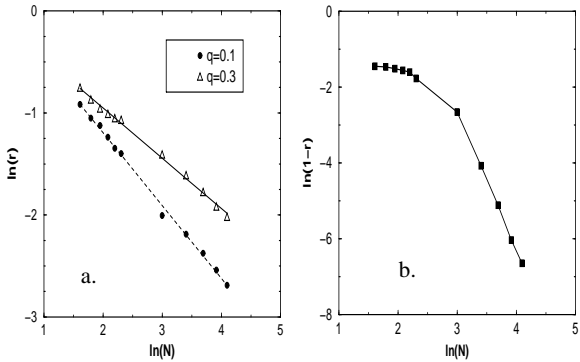,width=3.0in,angle=-0} \caption{Finite-size
scaling (log-log plots) for (a.) $q=0.1$ and $q=0.3$ (b.) $q=0.7$.
The best-fit lines for (a.) have slopes of $-0.714$ and $-0.4928$,
respectively.} \label{fig6}
\end{figure}

\section{A more general case}

Next, we briefly present our results for a more general case,
where the $S_i$ factors are also randomly distributed.

\begin{figure}[htp]
\epsfig{figure=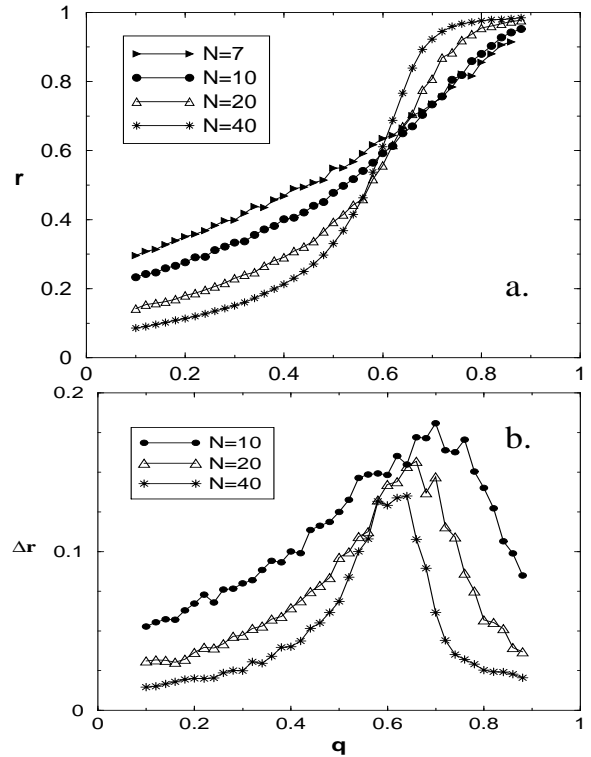,width=3.0in, height=4.0in, angle=-0}
\caption{Simulated annealing results for uniformly distributed
$S_i$ values. (a.) variation of the order parameter, (b.) standard
deviation of the order parameter as a function of $q$.}
\label{fig7}
\end{figure}

Considering a simple uniform
distribution of the $S_i$ values on the $[0,1]$ interval and $Z_{ij}$ distributed
according to the (\ref{prob}) distribution, we performed
a simulated annealing optimization. Since the variance of $J_{ij}$ scales again
like $N^{-2}$, the transition is naturally expected.
For $N=10,20$ and $40$ results supporting this geometrical phase transition
are plotted on Fig.~7.

\section{Discussions}

The observed geometrical phase transition is interesting also from the viewpoint
of the much discussed social percolation \cite{solomon}, where the
emergence of a giant cluster is observed in many social phenomena.
Our simple model suggests that large sociological systems can show tendencies to
percolation-like behavior due to coalition formation phenomena.
If a globally coupled large system has more propensities pointing towards collaboration
than conflict, usually a single coalition satisfies optimally the
apparently conflicting interactions. Contrary, when there are more conflicting propensities
than collaborative ones, the society will fragment in large number of
coalitions, and each element will isolate itself from the others.
As expected, this percolation-like behavior is rather smooth for small
systems sizes. The observed percolation-like behavior is also quite stable
relative to the choice of the $Z_{ij}$ propensities and  $S_i$ weight-factors.

It is also important to mention that according to the considered model the
most unpredictable societies are the "equilibrated" ones, where the
number of positive and negative links are roughly the same. From our numerical
results one can see that in this case $\Delta r$ is big, and the value
of $r$ is changing strongly with small variations of $q$. First, this means that the system
is very sensitive to the explicit realization of the $J_{ij}$ values. Secondly,
as seen in Fig.~4 in this region many equilibrium configurations with different $r$ values might co-exist,
all of them having the same minimal $K$ value (degeneracy of the ground-state might be high).
Third, a small difference in the measured
$q$ value can result in large differences for the expected $r$ values.
In these "equilibrated" societies statistical methods are useless for predicting the
optimal clusterization. Specific analyses of the concrete situation is thus the
only acceptable prediction method.

The fact that in the ground-state many equally-optimum configurations
with quite different maximal cluster sizes are possible might also
lead to interesting implications. It might well be possible the existence of some
"mixed" states, where the system behavior can be described not from a clear
coalition structure, but rather from a superposition of many coalition structures.

The model considered by us is of course a very simple one, capturing only a few parameters
that are important in understanding social coalition formation.
In our model we have also neglected the dynamics of the system, and
presumed that the system will clusterize in one of the optimal configurations.
The system is however frustrated, and many configurations with local minimum exist.
During its dynamics, the system might get trapped in a local minimum, and the
formed coalitions might be the one corresponding to this case, rather than the
global optimum case. The model considered by us and our results are usable thus
only for statistically predicting the optimal clusterization, and not for understanding the
coalitions that are formed in reality.

In conclusion, in the present study we presented evidences for a geometrical phase
transition in the ground-state of an infinite-range Potts glass where the
standard deviation of the bonds scale as $N^{-2}$.
For finite system sizes three
different methods were used to approach this NP hard optimization problem, all of
them supporting the percolation-like behavior of the largest cluster size
as a function of the positive links in the system. The model considered by us might be
useful in understanding some social percolation phenomena in large sociological systems.

\section{Acknowledgments}

The present study was sponsored by the Sapientia Foundation in Cluj.
We also thank the Bergen Computational Physics Laboratory in the framework of the
European Community - Access to Research Infrastructure action of the Improving Human
Potential Programme for supporting and financing our computations.


\end{document}